\documentclass[
prb,showpacs,superscriptaddress,numerical,amsmath,amssymb,floatfix,
reprint
]{revtex4-2}
\usepackage{xcolor}
\usepackage[
pdffitwindow=true,
colorlinks=true,
frenchlinks=false,
linkcolor=blue,
anchorcolor=blue,
citecolor=blue,
filecolor=blue,
urlcolor=blue,
bookmarks=true,
bookmarksopen=true,
bookmarksnumbered=true,
bookmarksopenlevel=1,
plainpages=false,
pdfpagelabels=true,
breaklinks
]{hyperref}
\usepackage[per-mode=symbol,separate-uncertainty]{siunitx}
\usepackage{graphicx}
\usepackage{dcolumn}
\usepackage{bm}
\usepackage[english]{babel}
\usepackage{color}
\usepackage{todonotes}
\usepackage{hyperref}
\usepackage{units}

\begin{document}
\title{Influence of low-energy magnons on magnon Hanle experiments in easy-plane antiferromagnets}

\author{Janine~G{\"u}ckelhorn}
\email{janine.gueckelhorn@wmi.badw.de}
\affiliation{Walther-Mei{\ss}ner-Institut, Bayerische Akademie der Wissenschaften, D-85748 Garching, Germany}
\affiliation{Physik-Department, Technische Universit\"{a}t M\"{u}nchen, D-85748 Garching, Germany}
\author{Akashdeep Kamra}
\email{akashdeep.kamra@uam.es}
\affiliation{Condensed Matter Physics Center (IFIMAC) and Departamento de F\'{i}sica Te\'{o}rica de la Materia Condensada, Universidad Aut\'{o}noma de Madrid, E-28049 Madrid, Spain}
\author{Tobias~Wimmer}
\affiliation{Walther-Mei{\ss}ner-Institut, Bayerische Akademie der Wissenschaften, D-85748 Garching, Germany}
\affiliation{Physik-Department, Technische Universit\"{a}t M\"{u}nchen, D-85748 Garching, Germany}
\author{Matthias~Opel}
\affiliation{Walther-Mei{\ss}ner-Institut, Bayerische Akademie der Wissenschaften, D-85748 Garching, Germany}
\author{Stephan~Gepr{\"a}gs}
\affiliation{Walther-Mei{\ss}ner-Institut, Bayerische Akademie der Wissenschaften, D-85748 Garching, Germany}
\author{Rudolf~Gross}
\affiliation{Walther-Mei{\ss}ner-Institut, Bayerische Akademie der Wissenschaften, D-85748 Garching, Germany}
\affiliation{Physik-Department, Technische Universit\"{a}t M\"{u}nchen, D-85748 Garching, Germany}
\affiliation{Munich Center for Quantum Science and Technology (MCQST), D-80799 M\"{u}nchen, Germany}
\author{Hans~Huebl}
\affiliation{Walther-Mei{\ss}ner-Institut, Bayerische Akademie der Wissenschaften, D-85748 Garching, Germany}
\affiliation{Physik-Department, Technische Universit\"{a}t M\"{u}nchen, D-85748 Garching, Germany}
\affiliation{Munich Center for Quantum Science and Technology (MCQST), D-80799 M\"{u}nchen, Germany}
\author{Matthias~Althammer}
\email[]{matthias.althammer@wmi.badw.de}
\affiliation{Walther-Mei{\ss}ner-Institut, Bayerische Akademie der Wissenschaften, D-85748 Garching, Germany}
\affiliation{Physik-Department, Technische Universit\"{a}t M\"{u}nchen, D-85748 Garching, Germany}

\begin{abstract}
Antiferromagnetic materials host pairs of spin-up and spin-down magnons which can be described in terms of a magnonic pseudospin. The close analogy between this magnonic pseudospin systems and that of electronic charge carriers led to the prediction of fascinating phenomena in antiferromagnets. Recently, the associated dynamics of antiferromagnetic pseudospin has been experimentally demonstrated and, in particular, the first observation of the magnon Hanle effect has been reported. We here expand the magnonic spin transport description by explicitly taking into account contributions of finite-spin low-energy magnons. In our experiments we realize the spin injection and detection process by two Platinum strips and investigate the influence of the Pt-strips on the generation and diffusive transport of magnons in films of the antiferromagnetic insulator hematite. For both a $\SI{15}{\nano\meter}$ and a $\SI{100}{\nano\meter}$ thick film, we find a distinct signal caused by the magnon Hanle effect. However, the magnonic spin signal exhibits clear differences in both films. In contrast to the thin film, for the thicker one, we observe an oscillating behavior in the high magnetic field range as well as an additional offset signal in the low magnetic field regime. We attribute this offset signal to the presence of finite-spin low-energy magnons.
\end{abstract} 

%----------------------------------------------------------------------------------------------------%

\maketitle

%-----------------------------------------Intro------------------------------------------------------- %
\section{Introduction}\label{sec:intro}
\raggedbottom
Magnonic spin currents have become the active ingredient in an emerging paradigm for spin and information transport
via magnons, the elementary excitations of the spin system in magnetically ordered insulators. They have drawn much attention due to their potential applications in information processing at a low dissipation~\cite{Bauer2012,Chumak2015,CornelissenMMR,Nakata2017,Yuan2018,Althammer2018,Althammer2021,Klaui2018,Hou2019,Wimmer2019,Kamra2019}. In particular, antiferromagnetic insulators (AFIs) offer new opportunities for interesting device applications due to their immunity to stray fields~\cite{Jungwirth2016AFM,BaltzAFM2018}, their magnetic resonance frequencies in the terahertz regime~\cite{Jungwirth2016AFM,BaltzAFM2018,Li2020,Vaidya2020}, and ultrafast response times~\cite{Nemec2018,Olejnik2018}. Within the last decade, the magnetic dynamics and transport properties of antiferromagnetic insulators have been intensively studied~\cite{BaltzAFM2018,JungfleischAFM2018,Hou2019}. It was shown that heterostructures consisting of AFIs and heavy metals with strong spin-orbit coupling enable the study of magnon spin transport in AFIs via electronic injection and detection~\cite{CornelissenMMR,SchlitzMMR,Wimmer2019,CornelissenTheory,ZhangMMR1,Li2016}. In a two-strip configuration, the spin Hall effect (SHE) in two spatially separated heavy metal electrodes is utilized to inject and detect magnonic spin currents~\cite{HirschSHE,SaitohISHE,Sinova2015,Valenzuela}.
Micrometer long spin transport has first been observed in easy-axis AFIs in similiar devices~\cite{Klaui2018}. Apart from SHE-induced magnons, the transport of thermally generated magnonic spin currents via the spin Seebeck effect has also been reported~\cite{Geert2020,Ross2021}. 

In these studies, a simple intuitive understanding can be accomplished by assuming a variation of the magnon chemical potential along only one direction~\cite{CornelissenMMR,CornelissenTheory,Wimmer2020,Kamra2020}. This is a good approximation in magnetic layers much thinner than the magnon diffusion length~\footnote{The magnon transport is diffusive in nature and captured by a spatial variation of its chemical potential. Hence, the magnon propagation or decay length has been called the magnon ``diffusion length'' in literature~\cite{CornelissenMMR,CornelissenTheory,SchlitzMMR}. This terminology is to be contrasted with the diffusion of conserved particles, such as a gas diffusing into a room.}. In a prototypical ferro(ferri)magnet - yttrium iron garnet (YIG), a numerical analysis capturing the variation of the magnon chemical potential across the thickness of the magnetic layer was found to be important for reproducing certain experimental features in films with thicknesses comparable to the magnon diffusion length~\cite{Shan2016}. The magnetic layers and the corresponding magnon density of states in this treatment are assumed 3-dimensional. No new features have been found in the magnon spin transport studies when the thickness of the YIG film becomes larger than the magnon thermal wavelength (typically much smaller than the magnon diffusion length), i.e., when the dimensionality of the magnetic layer changes from quasi 2D to quasi 3D. On the other hand, spin Seebeck effect measurements in YIG with different thicknesses provide evidence for an important role of the magnon density of states and low-energy magnons when crossing the magnon thermal wavelength boundary~\cite{Kehlberger2015}. This raises the question whether there are any magnon spin transport features that depend sensitively on the effective dimensionality of the magnetic layer. We aim to experimentally examine this question here, for easy-plane antiferromagnets.

In contrast to magnons in ferromagnetic insulators, which carry spin of only one direction, AFIs host pairs of spin-up and spin-down magnons as the eigenmodes and thus enable superpositions. This can result in linearly polarized oscillations of the N\'eel order forming zero-spin excitations~\cite{Kamra2017,Liensberger2019}. Due to the existence of this configuration in easy-plane AFIs, they were considered unlikely to transport angular momentum. This opened the question which type of magnonic transport are supported, by this particular configuration.
In the last year, the long-distance transport of SHE-induced magnonic spin currents has been experimentally addressed in easy-plane AFIs~\cite{Wimmer2020,Han2020,Lebrun2020,Ross2020}. 
On a theoretical level, these antiferromagnetic magnon pairs can be described as a pseudospin system in an analogous manner as the two-level spin system of electronic charge carriers~\cite{Cheng2016,Daniels2018,Shen2020,Kawano2019}. Due to this formal equivalence between electron spin and antiferromagnetic magnon pseudospin, analogous phenomena are predicted in AFIs as they occur in electron spin systems~\cite{Cheng2016,Daniels2018,Shen2020,Kawano2019,Cheng2016B,Zyuzin2016,Kawano2019B}.
In particular, the first observation of the magnonic analog of the electronic Hanle effect~\cite{Fabian2007,Kikkawa1999,Jedema2002} has been reported in thin films of the AFI hematite ($\alpha-\mathrm{Fe}_2\mathrm{O}_3$) in our previous letter~\cite{Wimmer2020}. This is achieved by the realization of a coherent control of the magnon spin and its transport in the AFI thin film. A similar magnon-based Hanle effect has been observed in Zn-doped hematite~\cite{Ross2020}. Besides these experimental observations, the antiferromagnetic pseudospin dynamics have also been investigated theoretically~\cite{Kamra2020,Shen2021}. 
Present studies focused on thin films and the experimental findings were described using a one-dimensional pseudospin transport model. 
As discussed above, the validity of this model encounters two important length scales associated with magnons - namely the thermal wavelength and the diffusion length - as the thickness of the magnetic film increases. The former scale determines the effective dimensionality of the magnet, thereby influencing the magnon density of states and their role in spin transport. Here, we wish to experimentally probe this transition characterized by the magnon thermal wavelength and appropriately supplement the theoretical model~\cite{Kamra2020} in order to account for the transition.
To this end, we investigate magnonic spin transport in hematite films with varying thickness at different temperatures. The theoretical model~\cite{Kamra2020} is accordingly expanded to incorporate the role of finite thickness and a direct evaluation of the experimentally measured magnetoresistance utilizing a two strip device.

In this article, we first present our theoretical model, which enables a clear discussion and interpretation of the experiments reported later. Hence, we begin by briefly summarizing the magnon pseudospin concept of spin transport in AFIs. Subsequently, we expand the model by accounting for low-energy magnons and explicitly taking the spin injection and detection into account. In the next part, we present our experimental results on magnon spin transport in films of hematite with varying thickness utilizing the two-strip configuration. The magnon spin signal measured at the detector exhibits an oscillation of the polarity as a function of the applied magnetic field, which we explain in terms of an antiferromagnetic magnon Hanle effect. Interestingly, for thicker hematite layers we find a large positive offset in the magnon spin signal. We attribute this to the typical diffusive transport of low-energy finite-spin magnons. We discuss our results within the scope of our expanded description of the magnon pseudospin dynamics.

%-----------------------------------------Theory------------------------------------------------------- %
\section{Theoretical Model}\label{sec:theory}

We begin with a discussion of the theoretical model describing magnonic spin transport in an AFI. This provides us the tools and terminology for a clear discussion and interpretation of our experiments reported further below. A detailed theoretical analysis of magnon transport in AFIs with arbitrary anisotropies has been detailed elsewhere~\cite{Kamra2020}. Here, we first repeat some key elements of this magnon pseudospin-based transport theory~\cite{Kamra2020} (Sec.~\ref{sec:pseudotransport}). Then, we augment this description by accounting for the contribution of finite-spin low-energy magnons (Sec.~\ref{sec:finitespin}), which appear to underlie certain aspects of our experiments on thick AFI films. 
In the following, we will differentiate between thin and thick AFI films. As we are interested here in characterizing the change that accompanies the magnetic film going from quasi-2D to quasi-3D, we compare our film thickness to the thermal magnon wavelength. The latter is typically much smaller than the magnon diffusion length, and we work under this assumption. We refer to thick films when the film thickness $t_\mathrm{m}$ is on the order or larger than the magnon thermal wavelength $l_\mathrm{th}$, while for thin films $t_\mathrm{m} \ll l_\mathrm{th}$ applies.
Finally, we consider the spin injection and detection process explicitly thereby evaluating the experimentally measured magnetoresistance (Sec.~\ref{sec:InjDet}). We consider a simplified and analytically tractable theoretical model, which is, strictly speaking, not valid in the full range of our experiments. Nevertheless, it allows us to understand all the qualitative features, while providing insights on the reasons for deviations of theory from the experiments in certain regimes.

%----------------------------------Magnon Pseudospin Transport----------------------------------------------- %
\subsection{Magnon pseudospin transport}\label{sec:pseudotransport}

\begin{figure}[tb]
	\begin{center}
		\includegraphics[width=85mm]{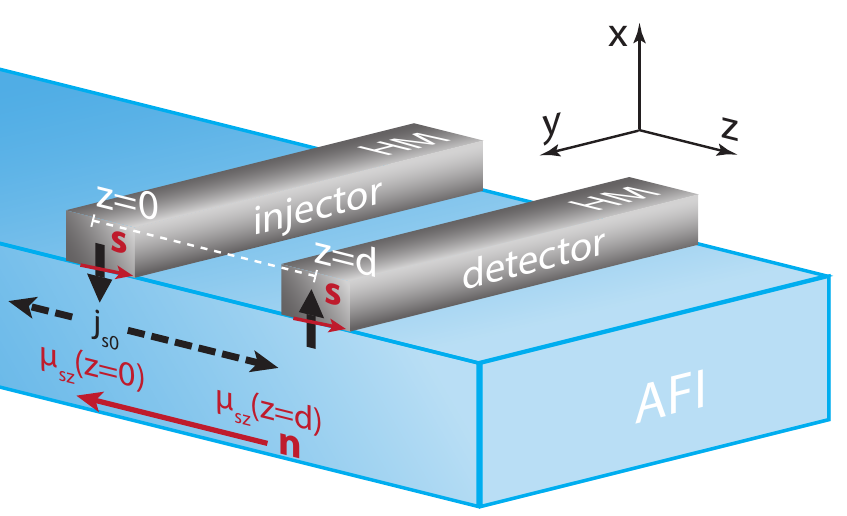}
		\caption{Device schematic for non-local magnon spin transport experiments. A $z$-polarized magnon spin and pseudospin currents are injected into and detected from the antiferromagnetic insulator (AFI) using two spatially separated heavy metal (HM) electrodes. The AFI is considered thin along the $x$-direction and infinite along the $z$-direction.}
		\label{fig:schematic}
	\end{center}
\end{figure}

We consider a device as depicted in Fig.~\ref{fig:schematic}. A charge current driven through the injector strip consisting of a heavy metal generates an electronic spin accumulation at the interface via the spin Hall effect~\cite{Sinova2015,HirschSHE}. This in turn injects magnonic spin current into the AFI which diffusively propagates towards another heavy metal electrode~\cite{CornelissenMMR,Klaui2018}. At this point, the magnon spin is detected as a voltage in this detector electrode via the inverse spin Hall effect. In this subsection, we focus on describing the spin transport in the AFI leaving a detailed discussion of the injection and detection processes to Sec~\ref{sec:InjDet}. We largely reproduce the key elements of the required spin transport theory that has been detailed elsewhere~\cite{Kamra2020}.

Two-sublattice AFIs admit a broad range of excitations formed from a superposition of spin-up and spin-down magnons as the basis states~\cite{Kamra2020,Wimmer2020,Kawano2019,Daniels2018,Cheng2016}. As a result, the magnonic excitations are described via a pseudospin vector on the unit Bloch sphere~\cite{Kamra2020,Wimmer2020}. The actual magnonic spin is given by the $z$-component of the pseudospin vector and is parallel to the equilibrium N\'eel order $\pmb{n}$. For example, the spin-1 magnons have their pseudospin pointing along $\pm \hat{\pmb{z}}$ and correspondingly carry unit spin. In the classical Landau-Lifshitz description, they correspond to spin waves with circularly precessing N\'eel vector. Magnons with spin 0 have their pseudospin lying in the $xy$-plane, and correspond to a linearly oscillating N\'eel vector. The nature of magnonic eigenmodes in the AFI is determined by the free energy landscape, e.g., magnetocrystalline anisotropies. For example, easy-axis AFIs have spin-1 magnonic eigenmodes while easy-plane AFIs host spin-0 modes. However, in nonequilibrium situations, any magnonic modes can exist in any AFI~\cite{Kamra2020}. 

The general description of magnon-mediated spin transport is conveniently captured via the dynamics and spatial evolution of the pseudospin chemical potential $\pmb{\mu}_\mathrm{s}$~\cite{Kamra2020}:
\begin{eqnarray}\label{eq:pseudodiff}
\frac{\partial \pmb{\mu}_\mathrm{s}}{\partial t} & = & D_\mathrm{m} \nabla^2 \pmb{\mu}_\mathrm{s} - \frac{\pmb{\mu}_\mathrm{s}}{\tau_\mathrm{m}} + \pmb{\mu}_\mathrm{s} \times \pmb{\omega},
\end{eqnarray}
where $D_\mathrm{m}$ is the magnon diffusion constant, and $\pmb{\omega}$ is the pseudofield. The second term on the right hand side accounts for magnon losses at a rate $\tau_{\mathrm{m}}^{-1}$ and the last term a pseudospin torque appearing when $\pmb{\mu}_\mathrm{s}$ is not parallel to $\pmb{\omega}$. In contrast to the usual scalar-valued magnon chemical potential for ferromagnetic systems, the pseudospin chemical potential is a vectorial quantity, which thereby accounts for the eigenmode information. Here, $\pmb{\omega}$ absorbs the material details, such as anisotropies, and allows for an ensemble averaged description of spin transport. It also determines the magnonic eigenmodes, since their pseudospin vector is collinear with $\pmb{\omega}$. We consider $\pmb{\omega} = \omega_x \hat{\pmb{x}} + \omega_z \hat{\pmb{z}}$ as it allows us to capture the complete range of magnonic eigenmodes from spin-1 ($\omega_x = 0$) to spin-0 ($\omega_z = 0$, $\omega_x\neq 0$). The injection of spin into the AFI is taken into account via the boundary conditions.

Solving Eq.~\eqref{eq:pseudodiff} in three-dimensions, i.e. for thick films, is analytically intractable. Hence, we consider a thin film such that $\pmb{\mu}_\mathrm{s}$ does not depend on $x$ (cf. Fig.~\ref{fig:schematic}). We further consider the system to be translationally invariant along the $y$-direction such that $\pmb{\mu}_\mathrm{s}$ only depends on $z$. Thus, in steady state  Eq.~\eqref{eq:pseudodiff} simplifies to:
\begin{align}
D_\mathrm{m} \frac{\partial^2 \pmb{\mu}_\mathrm{s}}{\partial z^2}  - \frac{\pmb{\mu}_\mathrm{s}}{\tau_\mathrm{m}} + \pmb{\mu}_\mathrm{s} \times \left( \omega_x \hat{\pmb{x}} + \omega_z \hat{\pmb{z}} \right) & = 0,
\end{align}
with the boundary conditions accounting for magnon spin injection:
\begin{eqnarray}
- \left. D_\mathrm{m} \chi \frac{\partial \mu_{\mathrm{s}z}}{\partial z} \right|_{z=0} & = & j_{\mathrm{s}0}, \label{eq:bc1} \\
\left. \frac{\partial \mu_{\mathrm{s}x,\mathrm{s}y}}{\partial z} \right|_{z=0} & = & 0. \label{eq:bc2}
\end{eqnarray}
Here, $j_{\mathrm{s}0}$ is the magnon spin current density that flows along the positive $z$-direction at the injector location $z = 0$. This will be discussed further in Sec.~\ref{sec:InjDet} below. $\chi$ is the susceptibility that relates the total magnon pseudospin density $\pmb{\mathcal{S}}$ to the pseudofield $\pmb{\omega}$ via $\pmb{\mathcal{S}} = \chi (\pmb{\omega} + \pmb{\mu}_\mathrm{s})$~\cite{Kamra2020}. In addition, we assume $\pmb{\mu}_\mathrm{s}(z \to \infty) = 0$ consistent with the requirement that injected spin decays at large distances. After some algebra, the solution for $\mu_{\mathrm{s}z}(z)$ is obtained as~\cite{Kamra2020}:
\begin{eqnarray}
\mu_{\mathrm{s}z} (z) & = & \mu_{\mathrm{osc}}(z) + \mu_{\mathrm{dec}}(z),  \label{eq:solbeg}\\
\mu_{\mathrm{osc}}(z) & = & \frac{\omega_x^2}{\omega_x^2 + \omega_z^2} \ \frac{l_\mathrm{m} j_{\mathrm{s}0}}{D_\mathrm{m} \chi \left(a^2 + b^2\right)} \ e^{- \frac{a z}{l_\mathrm{m}}} \nonumber\\
&\times& \left[ -b \sin \left( \frac{b z}{l_\mathrm{m}} \right) + a \cos \left( \frac{b z}{l_\mathrm{m}} \right) \right], \label{eq:osc}\\
\mu_{\mathrm{dec}}(z) & = & \frac{\omega_z^2}{\omega_x^2 + \omega_z^2} \ \frac{l_\mathrm{m} j_{\mathrm{s}0}}{D_\mathrm{m} \chi}  \ e^{- \frac{z}{l_\mathrm{m}}}, \label{eq:dec}
\end{eqnarray}
where $l_\mathrm{m} \equiv \sqrt{D_\mathrm{m} \tau_\mathrm{m}}$ is the magnon diffusion length and we have additionally defined
\begin{align}
a \equiv & \frac{1}{\sqrt{2}} \sqrt{1 + \sqrt{1 + \beta^2}}, \label{eq:a} \\
b \equiv & \frac{1}{\sqrt{2}} \sqrt{- 1 + \sqrt{1 + \beta^2}}, \\
\beta^2  \equiv &  \tau_\mathrm{m}^2 \left(\omega_{x}^2 + \omega_z^2  \right). \label{eq:solend}
\end{align}
For a weakly coupled detector, as discussed further below, the detected magnon spin signal at position $z$ is proportional to $\mu_{\mathrm{s}z}(z)$. Thus, Eqs.~\eqref{eq:solbeg}-\eqref{eq:solend} describe the magnonic spin transport in the AFI.

We see that the spin propagation in the AFI consists of two distinct contributions. The first, described by Eq.~\eqref{eq:dec}, results from the magnonic eigenmodes bearing a finite spin or equivalently, $z$-projection of the pseudospin. This is the only mode of spin transport in easy-axis AFIs that host spin-1 magnons as eigenexcitation~\cite{Klaui2018,Klaui2020}. If the AFI eigenmodes would be perfectly spinless, i.e., bear pseudospin along the $x$-axis ($\omega_z = 0, \omega_x \neq 0$) in the considered case, this contribution would be absent. However, in the easy-plane phase, e.g. present in hematite investigated here, a small but finite anisotropy within the easy-plane~\cite{Lebrun2020,Wang2021,Boventer2021}, gives rise to a correspondingly small but finite spin of the magnonic eigenmodes (corresponding to pseudofield components $|\omega_x|\gg|\omega_z|>0$). We call this contribution the ``finite-spin signal'' and note that it decays with the usual magnon spin diffusion length $l_\mathrm{m}$. 

The second contribution to spin transport [Eq.~\eqref{eq:osc}] stems from the oscillation of magnon spin with time. Since the spin-1 magnons injected by the heavy metal do not correspond to the eigenmodes when $\omega_x \neq 0$, their properties evolve with time as captured by pseudospin precession about the pseudofield [Eq.~\eqref{eq:pseudodiff}]~\cite{Wimmer2020}. The corresponding precession frequency $\omega$ depends on various contributions (e.g., anisotropy, Dzyaloshinskii-Moriya interaction) to the magnetic free energy density and can generally be tuned via an applied magnetic field. This dependence of $\omega \approx \omega_x$ in hematite provides an experimental handle to control the magnon spin reaching the detector electrode~\cite{Wimmer2020}. Thus, the detected spin signal manifests oscillations characteristic of the Hanle effect with $\omega_x$ vanishing at a finite compensation field $H_\mathrm{c}$. We call this contribution the ``Hanle signal'' and note that it decays faster than $l_\mathrm{m}$ [Eq.~\eqref{eq:osc}]. The reason for this faster decay is that magnons take different trajectories from the injector to the detector thereby arriving with different phases and interfering destructively~\cite{Fabian2007,Jedema2002,Kamra2020}. 

Unlike the case of electrons, in which the spin precession and Hanle effect are rooted in the Zeeman coupling of the electron spin to the applied magnetic field~\cite{Fabian2007,Jedema2002,Kikkawa1999}, the pseudofield $\pmb{\omega}$ is determined by various free energy contributions describing the AFI and is model dependent. Two related~\cite{Wimmer2020,Ross2020,Shen2021}, but different, origins have been suggested to dominate $H_\mathrm{c}$. Here, we treat it as an experimentally observed field and perform a Taylor expansion of $\omega \approx \omega_x$ around $H_\mathrm{c}$ in analyzing our data.

%---------------------------------Finite-spin contribution without pseudospin precession--------------------------- %
\subsection{Finite-spin contribution from low-energy magnons}\label{sec:finitespin}

\begin{figure}[tb]
	\begin{center}
		\includegraphics[width=85mm]{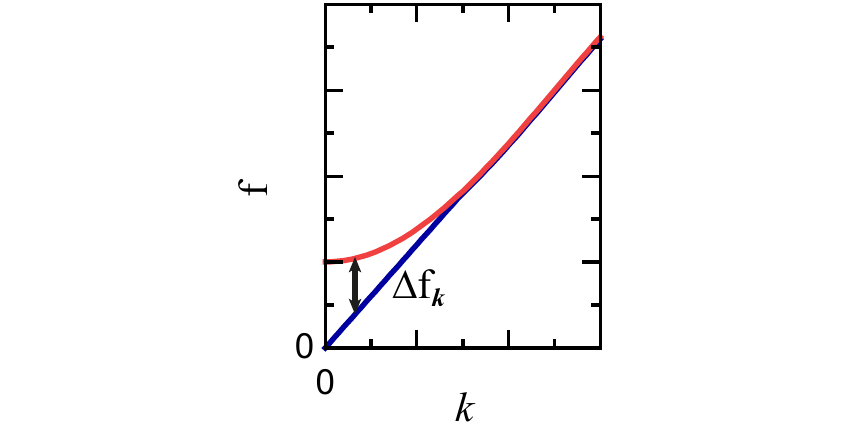}
		\caption{Schematic depiction of the low-$k$ magnonic dispersion of the two eigenmodes at zero applied magnetic field in hematite~\cite{Shen2021}. The lower branch contributes to the magnon transport due to the finite spin of the corresponding magnons.}
		\label{fig:dispersion}
	\end{center}
\end{figure}

The magnon pseudospin chemical potential $\pmb{\mu}_\mathrm{s}$ adequately describes the magnonic spin transport in nearly all of the Brillouin zone~\cite{Kamra2020}. However, it is unable to capture the role of low-energy magnons on account of an assumption discussed further below. The former are disregarded very often as they constitute a small fraction of the total magnons in the system~\cite{CornelissenTheory}. However, they have much longer scattering lengths and are excited strongly by the injector due to their high pre-existing thermal occupation~\cite{ZhangMMR1,CornelissenTheory,Zhang2012}. As a result, many experiments provided evidence for an important and special role of these low-energy magnons~\cite{Wimmer2019,Kikkawa2015,Kehlberger2015}. We now qualitatively discuss their role in the magnon spin transport in AFIs.

The pseudospin description validity at a given wavevector $\pmb{k}$ assumes that the frequency difference $\Delta f_{\pmb{k}}$ between the two eigenmodes is much smaller than their average frequency~\cite{Kamra2020}. If these two quantities start to be comparable, initially we merely lose the validity of linear response while the qualitative physics is still described adequately by the pseudospin picture. When the situation goes to an extreme with increasing $\Delta f_{\pmb{k}}$, as is the case for the lower magnon branch around $k = 0$ in hematite at zero applied field [see Fig.~\ref{fig:dispersion}]~\cite{Shen2021}, the pseudospin chemical potential completely fails to capture their contribution to the spin transport. Luckily, the pseudospin remains a well-defined and useful quantity. Its $z$-projection still corresponds to the spin carried by the eigenmodes~\cite{Kamra2020}. Thus, these low-energy magnons simply contribute to the finite-spin signal, without contributing to the pseudospin precession. Their contribution can hence be absorbed into Eq.~\eqref{eq:dec} as another offset that, in principle, would decay on a longer length scale than $l_\mathrm{m}$.

The role of this ad hoc finite-spin contribution from low-energy magnons is expected to diminish as an applied magnetic field increases their energy~\cite{Shen2021} and they start to be adequately described via the pseudospin chemical potential. Furthermore, in very thin films, the density of low-energy magnons is reduced considerably as the boundary condition along the film thickness imposes a finite and large $k$, thereby effectively gapping the low-energy magnons out. Thus, this contribution is expected to be relevant only in thick AFI films at low applied magnetic fields.

%----------------------------------Injection and detection----------------------------------------------- %
\subsection{Spin injection and detection}\label{sec:InjDet}

\begin{figure}[tb]
	\begin{center}
		\includegraphics[width=85mm]{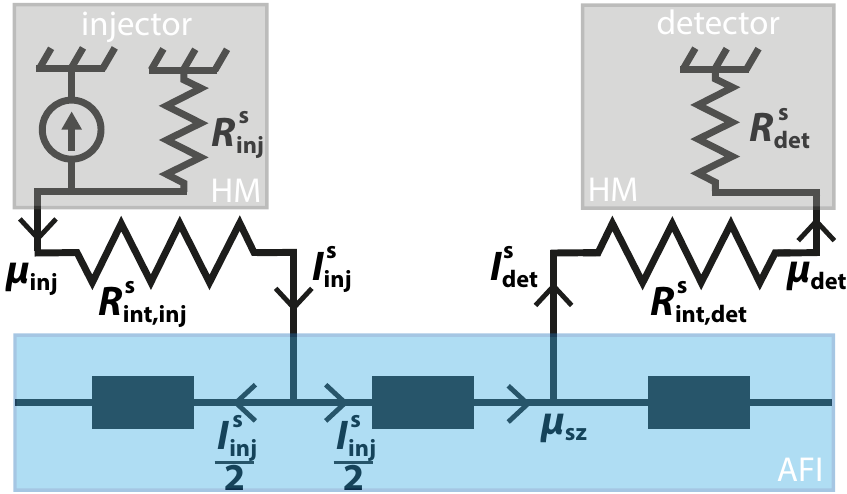}
		\caption{Equivalent circuit diagram for magnonic spin injection and detection. A current driven through the injector lead generates a spin accumulation $\mu_{\mathrm{inj}}$ in the metal. This, in turn, injects a spin current $I^\mathrm{s}_{\mathrm{inj}} \approx \mu_{\mathrm{inj}} / R^\mathrm{s}_{\mathrm{int,inj}}$ into the AFI which propagates and evolves as pseudospin transport, not described via a simple resistor. The detected spin current $I^\mathrm{s}_{\mathrm{det}} \approx \mu_{\mathrm{s}z}(z) / R^\mathrm{s}_{\mathrm{int,det}}$ senses the magnon spin chemical potential at the detector location.}
		\label{fig:circuit}
	\end{center}
\end{figure}

In our discussion above, we focused on the pseudospin and spin transport in the AFI. We treated magnon spin injection by assuming a $z$-polarized pseudospin current density $j_{\mathrm{s}0}$ at $z= 0$, and $\mu_{\mathrm{s}z}$ was treated as the detected magnon spin signal. This is indeed sufficient for understanding the spin propagation and dynamics in the AFI. However, a direct relation between the charge current driven in the injector and the voltage registered by the detector is not obtained in this approach. We take up this task in the present section.

A complete analysis of the problem at hand involves several parameters and becomes tedious, even for the much simpler case of ferromagnets~\cite{CornelissenTheory}. Hence, in order to model the essential physics, we work in the approximation that the AFI is weakly coupled to the injector and detector leads. An equivalent circuit diagram describing spin flow in the system at hand is depicted in Fig.~\ref{fig:circuit}.

A charge current density $j_\mathrm{ci}~ \hat{\pmb{y}}$ driven through the injector generates a $z$-polarized electronic spin accumulation at its interface with the AFI~\cite{CornelissenTheory}:
\begin{align}\label{eq:kappai}
\mu_{\mathrm{inj}} & = 2 e \theta_\mathrm{i} l_\mathrm{si} \rho_\mathrm{i} \tanh \left( \frac{t_\mathrm{i}}{2 l_\mathrm{si}} \right) j_\mathrm{ci} \equiv \kappa_\mathrm{i} j_{ci} ,
\end{align}      
where $e$ is the elementary charge, $\theta_\mathrm{i}$ the spin Hall angle, $l_\mathrm{si}$ the spin diffusion length in the injector, $\rho_\mathrm{i}$ its resistivity, and $t_\mathrm{i}$ is the injector thickness. Considering the interfacial spin conductivity $g_\mathrm{i}$, the injected magnon spin current is given as~\cite{CornelissenTheory}:
\begin{align}\label{eq:Iinj}
I^{\mathrm{s}}_{\mathrm{inj}} & = g_\mathrm{i} w_\mathrm{i} L \left( \mu_{\mathrm{inj}} - \mu_{\mathrm{s}z} \right) \approx g_\mathrm{i} w_\mathrm{i} L \mu_{\mathrm{inj}}, 
\end{align}   
where $w_\mathrm{i}$ is the injector width, $L$ is the device length (along $\hat{\pmb{y}}$) assumed to be the same for injector, detector, and AFI. The approximation above [Eq.~\eqref{eq:Iinj}] is valid in the limit of $g_\mathrm{i} w_\mathrm{i} L \to 0$ such that the entire ``potential'' drops across the interface. Considering that only half of the injected spin current is directed towards the detector (positive $z$-direction), we obtain:
\begin{align}\label{eq:js0}
j_{\mathrm{s}0} & = \frac{g_\mathrm{i} \kappa_\mathrm{i}}{2 t_\mathrm{m} t_\mathrm{i}} I_{\mathrm{inj}}, 
\end{align}
where $t_\mathrm{m}$ is the AFI thickness and $I_{\mathrm{inj}} = j_\mathrm{ci} w_\mathrm{i} t_\mathrm{i}$ is the total charge current driven through the injector. Equation~\eqref{eq:js0} allows us to relate the assumed injected magnon spin current density [Eq.~\eqref{eq:bc1}] to the relevant experimental variable $I_{\mathrm{inj}}$.

In a similar fashion, the spin current injected into the detector electrode is obtained as~\cite{CornelissenTheory}:
\begin{align}\label{eq:Idet}
I^{\mathrm{s}}_{\mathrm{det}} & = g_\mathrm{d} w_\mathrm{d} L \left( \mu_{\mathrm{s}z}(z) - \mu_{\mathrm{det}} \right) \approx g_\mathrm{d} w_\mathrm{d} L \mu_{\mathrm{s}z}(z),
\end{align}
where $\mu_{\mathrm{s}z}(z)$ is the $z$-component of the magnon pseudospin chemical potential at the detector position $z$, $g_\mathrm{d}$ is the interfacial spin conductivity, and $w_\mathrm{d}$ is the detector width. With the interfacial spin current given by Eq.~\eqref{eq:Idet}, we evaluate the inverse spin Hall effect voltage generated under open circuit conditions as~\cite{Mosendz2010}:
\begin{align}\label{eq:Vdet}
V_{\mathrm{det}}^{\mathrm{el}} & = \frac{\kappa_\mathrm{d} g_\mathrm{d} L}{\hbar t_\mathrm{d}} \mu_{\mathrm{s}z}(z), 
\end{align}
where $\kappa_\mathrm{d}$ is defined similar to $\kappa_\mathrm{i}$ in Eq.~\eqref{eq:kappai}, but for the detector electrode. We may express Eq.~\eqref{eq:solbeg} in a dimensionless form:
\begin{align}\label{eq:musznorm}
\mu_{\mathrm{s}z}(z) & = \frac{j_{\mathrm{s}0} l_\mathrm{m}}{D_\mathrm{m} \chi} ~ \tilde{\mu}_{\mathrm{s}z}(z),
\end{align}
where the dimensionless variable $\tilde{\mu}_{\mathrm{s}z}(z)$ contains all the information about magnon pseudospin transport and dynamics in the AFI. 

Finally, employing Eqs.~\eqref{eq:js0}, \eqref{eq:Vdet}, and \eqref{eq:musznorm}, we obtain the experimentally measured magnetoresistance at the detector:
\begin{align}
\Delta R^\mathrm{el}_\mathrm{det} & = \frac{V_{\mathrm{det}}^{\mathrm{el}}}{I_{\mathrm{inj}}} = \frac{2 L}{\hbar} \ \frac{\kappa_\mathrm{d} g_\mathrm{d}}{2 t_\mathrm{d}} \ \frac{\kappa_\mathrm{i} g_\mathrm{i}}{2 t_\mathrm{i}} \frac{l_\mathrm{m}}{D_\mathrm{m} \chi t_\mathrm{m}} \ \tilde{\mu}_{\mathrm{s}z}(z) \\
  & = \frac{2 L}{\hbar} \ \frac{\kappa_\mathrm{d} \kappa_\mathrm{i}}{4 t_\mathrm{i} t_\mathrm{d}} \ \frac{l_\mathrm{m}}{D_\mathrm{m} t_\mathrm{m}} \ \left( \frac{g_\mathrm{i} g_\mathrm{d}}{\chi} \right) \ \tilde{\mu}_{\mathrm{s}z}(z). \label{eq:Rnl}
\end{align}
Thus, within our model, the injector-detector separation ($z$) dependence of the magnetoresistance directly probes the magnonic spin transport as given by $\tilde{\mu}_{\mathrm{s}z}(z)$. Its magnitude, however, depends on several factors involving the properties of injector, detector, AFI and their interfaces. Considering the temperature dependence of $\Delta R^\mathrm{el}_\mathrm{det}$, we expect the factor in brackets [Eq.~\eqref{eq:Rnl}] to be the dominant. All three quantities $g_\mathrm{i}$, $g_\mathrm{d}$, and $\chi$ are obtained by summing over all magnon modes and increase similarly as the number of magnons in the AFI increases with temperature~\cite{CornelissenTheory}. As a result, the term in brackets also increases with temperature. The temperature dependence of the remaining parameters in Eq.~\eqref{eq:Rnl} has been found to be weak in spin Hall magnetoresistance experiments~\cite{Althammer2013,Meyer2014}.

Our assumption of injector and detector electrodes being weakly coupled to the AFI allows us to derive the relatively simple result Eq.~\eqref{eq:Rnl} for the experimentally observed magnetoresistance $\Delta R^\mathrm{el}_\mathrm{det}$. Within this model, the absolute magnitude of $\Delta R^\mathrm{el}_\mathrm{det}$ depends on several details but its qualitative variation with injector-detector distance or applied field are directly given by the magnon transport in the AFI via $\tilde{\mu}_{\mathrm{s}z}(z)$. In experiments, the requirement of a detectable signal demands a not too weak coupling and thus our assumed model may not be very precise. In particular, deviations from our model are expected when the injector-detector distance is comparable to the magnon diffusion length $l_\mathrm{m}$. In that limit, the boundary conditions at the detector, in addition to those at the injector [Eqs.~\eqref{eq:bc1} and \eqref{eq:bc2}], need to be accounted for directly when considering the magnon transport in the AFI~\cite{CornelissenTheory}. Nevertheless, our simple analytic model succeeds in accounting for most features of the observed magnetoresistance in a wide range of devices.

%----------------------- Experimental Data ----------------------------------%
\section{Experiments}\label{sec:experiment}

Next, we present our experimental results on the magnon transport in the AFI hematite ($\mathrm{\alpha}$-$\mathrm{Fe}_2\mathrm{O}_3$) and discuss them in the context of our theoretical model introduced in Sec.~\ref{sec:theory}. In particular, we investigate the influence of dimensionality on the excited magnon spin signal. First, we describe the experimental details of the sample fabrication and the measurement methods (Sec.~\ref{sec:details}). Then, we discuss the magnon spin signal for a thin hematite film ($t_\mathrm{m}=\SI{15}{\nano\meter}$), for which we find similar results as in our previous work~\cite{Wimmer2020}. Here, we particulary investigate the influence of different injector-detector distances (Sec.~\ref{sec:15nm_film}). In addition, we investigate the magnon transport through a $\SI{100}{\nano\meter}$ thick hematite layer and observe significant changes in the magnon spin signal (Sec.~\ref{sec:100nm_film}) as compared to the thin film.
%------------------- Sample configuration --------------------------------------%
\subsection{Hematite films and measurement methods}\label{sec:details}

\begin{figure}[tb]
	\begin{center}
		\includegraphics[width=85mm]{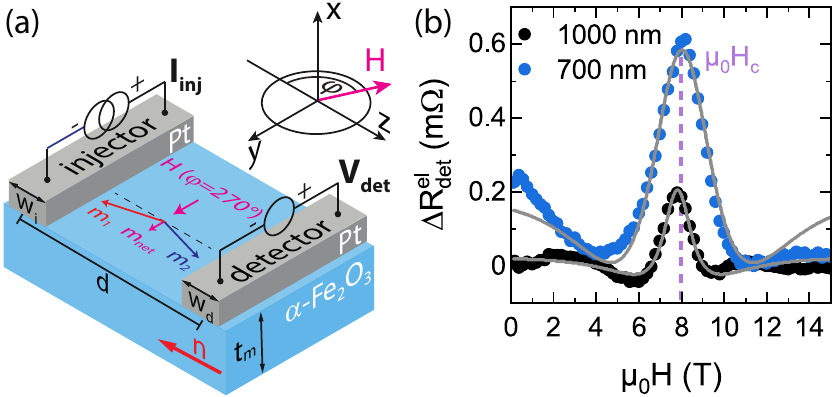}
			\caption{(a) Sketch of the sample configuration, the electrical wiring scheme, and the coordinate system with the in-plane rotation angle $\varphi$ of the applied magnetic field $\mu_0\pmb{H}$. The canting of the magnetic sublattices $\pmb{m}_\mathrm{1}$ and $\pmb{m}_\mathrm{2}$ is illustrated. The corresponding net moment $\pmb{m}_\mathrm{net}$ is aligned along the applied magnetic field $\mu_0\pmb{H}$, while the Néel order parameter $\pmb{n}$ is perpendicular to $\mu_0\pmb{H}$. (b) Amplitude of the electrically excited magnon spin signal $\Delta R_\mathrm{det}^\mathrm{el}$ as a function of the magnetic field strength for devices on a $t_m=\SI{15}{\nano\meter}$ thin film with a center-to-center distance of $d=\SI{1000}{\nano\meter}$ and $d=\SI{700}{\nano\meter}$, respectively, at $T=\SI{200}{\kelvin}$. The gray lines are fits using Eqs.~\eqref{eq:solbeg}-\eqref{eq:solend}.}
		\label{fig:fig1thinhematite}
	\end{center}
\end{figure}

The investigated hematite films are grown via pulsed laser deposition at the Walther-Meißner-Institut. Here, we deposit epitaxial $\mathrm{\alpha}$-$\mathrm{Fe}_2\mathrm{O}_3$ on (0001)-oriented $\mathrm{Al}_2\mathrm{O}_3$ substrates using a substrate temperature of $\SI{320}{\celsius}$, an oxygen pressure of $\SI{25}{\micro \bar}$, a laser fluence at the target of $\SI{2.5}{\joule\per\square\centi\meter}$ and a laser repetition rate of $\SI{2}{\hertz}$. The films have thicknesses of $t_\mathrm{m}=\SI{15}{\nano\meter}$ and $\SI{100}{\nano\meter}$. In the following, we refer to the first one as the thin film and the second as the thick film. Both films feature an out-of-plane Dzyaloshinskii-Moriya interaction (DMI) vector and an easy-plane phase over the whole temperature range due to the lack of the Morin transition as discussed in our previous work~\cite{Wimmer2020}. This means that the Néel vector $\pmb{n}$ and the two sublattice magnetizations $\pmb{m}_\mathrm{1}$ and $\pmb{m}_\mathrm{2}$ lie in  the (0001) plane or $yz$-plane [cf. Fig.\,\ref{fig:fig1thinhematite}(a)]. They are slightly canted due to the DMI, resulting in a net magnet moment $\pmb{m}_\mathrm{net}=\pmb{m}_\mathrm{1}+\pmb{m}_\mathrm{2}$, which is additionally controlled by the magnitude of the applied magnetic field. To investigate the magnon spin transport in the two thin films, we employ two narrow Pt strips as illustrated in Fig.\,\ref{fig:fig1thinhematite}(a). To this end, we use electron beam lithography to pattern two strip structures with varying center-to-center distances $d$ on top of the hematite films and deposit ex-situ polycrystalline $\SI{5}{\nano\meter}$ of Pt by magnetron sputtering. The Pt-strips of a device have either a width of $w_\mathrm{i}=w_\mathrm{d}=\SI{500}{\nano\meter}$ or $\SI{250}{\nano\meter}$ and a length of $L=\SI{100}{\micro\meter}$.
This strip configuration allows for an all-electrical generation and detection of pure spin currents.
 
For the experiments, we apply a DC charge current $I_\mathrm{inj}$ to the injector featuring typical current densities of $j_\mathrm{ci}\sim\SI{2e11}{\ampere\per\square\meter}$, which generates via the SHE an electron spin accumulation at the interface with the hematite. This spin accumulation leads to a finite spin and pseudospin magnon current in the AFI. Via the inverse SHE we electrically detect these magnons as a voltage signal $V_\mathrm{det}$ at the detector electrode.  The current reversal technique allows us to unambiguously assign the measured detector voltage $V_\mathrm{det}^\mathrm{el}=\left[ V_\mathrm{det}(+I_\mathrm{inj})-V_\mathrm{det}(-I_\mathrm{inj})\right]/2$ to the SHE-induced magnons transported from injector to detector~\cite{Ganzhorn2016,SchlitzMMR}. In order to account for the different geometries and injector current $I_\mathrm{inj}$, the magnon spin signal is given by $ R_\mathrm{det}^\mathrm{el}= V_\mathrm{det}^\mathrm{el}/I_\mathrm{inj}$.
The field-dependent and angle-dependent measurements are performed in two different vector magnet cryostats.

%--------------------------- Ultra thin film --------------------------------------%
\subsection{Thin hematite film}\label{sec:15nm_film}

First, we focus on the $\SI{15}{\nano\meter}$ thin hematite film. To investigate the magnetic field-dependence of the signal measured at the detector, we apply a magnetic field along $\hat{\pmb{y}}$, which orients the Néel vector $\pmb{n}$ along $-\hat{\pmb{z}}$. Only in this configuration, for $\pmb{H}\perp\pmb{n}$, we expect an electrical magnon excitation. In Fig.\,\ref{fig:fig1thinhematite}(b), the amplitude of the magnon spin signal $\Delta R_\mathrm{det}^\mathrm{el}$ is plotted as a function of the magnetic field strength $\mu_0H$ for two different injector-detector distances $d=\SI{1000}{\nano\meter}$ (black data points) and $d=\SI{700}{\nano\meter}$ (blue data points) at $T=\SI{200}{\kelvin}$. 

We observe a pronounced peak in the positive magnon spin signal regime at $\mu_0H \approx\SI{8}{\tesla}$ for both electrode spacings, which is in perfect agreement with our previous measurements~\cite{Wimmer2020}. For increasing or decreasing field strength, respectively, $\Delta R_\mathrm{det}^\mathrm{el}$ decreases until it approaches zero signal and exhibits a sign inversion, which is particularly pronounced for the structure with $d=\SI{1000}{\nano\meter}$. Comparing this with our theoretical model, which is described in detail in Sec.~\ref{sec:pseudotransport}, this behavior can be attributed to the ``Hanle signal''. 
The peak corresponds to the compensation field $\mu_0H_\mathrm{c}$ where the pseudofield vanishes $\omega=0$. Note that the peak position is independent of the electrode spacing $d$ as the condition for vanishing pseudofield is determined by the free energy landscape of the AFI. For $\Delta R_\mathrm{det}^\mathrm{el}=0$, the pseudospin density vector $\pmb{\mathcal{S}}$ is rotated by $\SI{90}{\degree}$, which means that the propagating magnon modes are linearly polarized and thus carry zero spin. The sign inverted spin signal corresponds to a $\SI{180}{\degree}$ rotation of the pseudospin vector and, therefore, an inversion of the magnon mode chirality. 
In contrast to the compensation field, the pseudofield remains finite ($\omega\neq0$) for the two latter cases and thus we expect that the magnetic field, where the magnon spin signal vanishes or exhibits a sign inversion, varies with the distance $d$. This is in agreement with our experimental data. 

Consistent with our model, we can describe the field-dependence of the magnon spin signal $\Delta R_\mathrm{det}^\mathrm{el}$ via Eqs.~\eqref{eq:solbeg}-\eqref{eq:solend}. The corresponding theoretical curves (gray lines) shown in Fig.~\ref{fig:fig1thinhematite}(b) reproduce well the measured data around the compensation field $\mu_0H_\mathrm{c}$. 
Our model is, strictly speaking, not valid in the full range of our experiments, which can be observed in the low and high field regime, respectively, where the theoretical curve and experimental data deviate. In particular, in the low field regime for $d=\SI{700}{\nano\meter}$ $\Delta R_\mathrm{det}^\mathrm{el}$ increases with decreasing field until it approaches a finite positive signal at $\mu_0H=\SI{0}{\tesla}$. 
This behavior is in contrast to the magnon spin signal from the structure with an electrode spacing of $d=\SI{1000}{\nano\meter}$, which approaches zero at zero field. As discussed above in Sec.~\ref{sec:InjDet}, deviations from the model are expected when the injector-detector distance is comparable to the magnon diffusion length, which is $l_\mathrm{m}\approx\SI{0.5}{\micro\meter}$ in our case~\cite{Wimmer2020}. While the black data points correspond to a structure with an edge-to-edge distance of $\SI{500}{\nano\meter}$, the blue data points are from a structure with an edge-to-edge distance of only $\SI{200}{\nano\meter}$, which is clearly smaller than $l_\mathrm{m}$. 
However, obtaining a full quantitative model to account for all these contributions is complex and not within the scope of this work.

%--------------------------- Thick film --------------------------------------%
\subsection{Thick hematite film}\label{sec:100nm_film}

\begin{figure}[tb]
	\begin{center}
		\includegraphics[width=85mm]{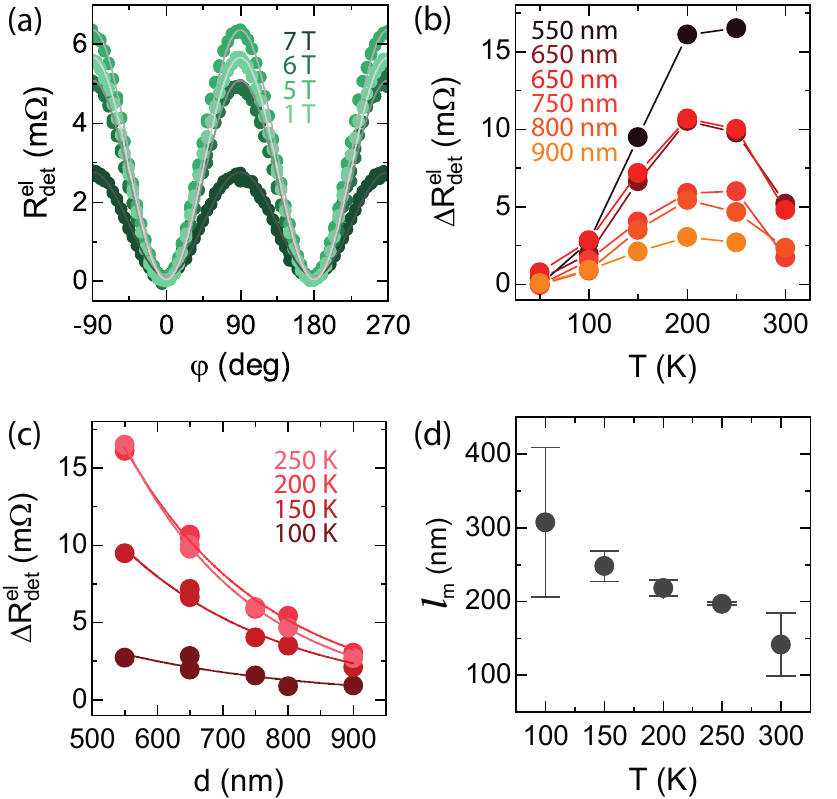}
		\caption{(a) Angle-dependent magnon spin signal $R_\mathrm{det}^\mathrm{el}\propto V_\mathrm{det}^\mathrm{el}/I_\mathrm{inj}$ of electrically induced magnons measured at the detector with an injector-detector distance of $d = \SI{800}{\nano\meter}$ for $T=\SI{200}{\kelvin}$ and various magnetic field magnitudes for the $\SI{100}{\nano\meter}$ thick hematite film. The solid gray lines are fits to a simple $\Delta R_\mathrm{det}^\mathrm{el}\sin^2(\varphi)$ function. (b) Electrically induced magnon spin signal amplitudes $\Delta R_\mathrm{det}^\mathrm{el}$, extracted from the angle-dependent measurements in (a), as a function of temperature $T$ for several injector-detector distances $d$. The solid lines are only guides to the eye. The data is measured at an external field strength of $\mu_0H = \SI{600}{\milli\tesla}$. (c) The amplitudes $\Delta R_\mathrm{det}^\mathrm{el}$ are plotted versus the distance $d$ for different temperatures $T$ at $\mu_0 H = \SI{600}{\milli\tesla}$. The solid lines are fits to Eq.~\eqref{eq:diffusionlength}. (d) The extracted magnon diffusion length $l_m$ is plotted as a function of $T$.}
		\label{fig:fig2MITcomparison}
	\end{center}
\end{figure}

Next, we discuss the results for the $\SI{100}{\nano\meter}$ thick film and compare them to those of the $\SI{15}{\nano\meter}$ thin sample as well as to our theoretical model. 
For a comprehensive study, we first measure the magnon spin signal $R_\mathrm{det}^\mathrm{el}$ as a function of the magnetic field orientation $\varphi$ as illustrated in Fig.~\ref{fig:fig1thinhematite}(a) with a fixed magnitude $\mu_0H$ at a temperature of $\SI{200}{\kelvin}$. The data is shown in Fig.~\ref{fig:fig2MITcomparison}(a) for a center-to-center distance $d=\SI{800}{\nano\meter}$. The results exhibit the distinctive $\sin^2(\varphi)$ angular variation expected for electrically induced diffusive transport from injector to detector~\cite{CornelissenMMR,SchlitzMMR}. As shown by the gray lines in Fig.~\ref{fig:fig2MITcomparison}(a) the angle dependence can be represented by a simple $\Delta R^\mathrm{el}_\mathrm{det}\sin^2(\varphi)$ function, where $\Delta R_\mathrm{det}^\mathrm{el}$ corresponds to the amplitude of the electrically induced magnon transport signal. Note that the electrical magnon excitation is largest when $\pmb{n}\parallel\pmb{\hat{z}}$, this means $\pmb{H}\parallel\pmb{\hat{y}}$ (as $\pmb{H}\perp\pmb{n}$) in our experiments. In accordance with previous experiments in AFIs, the magnon excitation originates from the antiferromagnetic Néel order and is shifted by $\SI{90}{\degree}$ compared to similar measurements in ferrimagnetic materials~\cite{CornelissenMMR,SchlitzMMR,Shan2017}. 
The quantity $\Delta R_\mathrm{det}^\mathrm{el}$ is extracted from the angle dependent measurements for different structures and plotted in Fig.~\ref{fig:fig2MITcomparison}(b) as a function of temperature in the range $T=50-\SI{300}{\kelvin}$ for different distances at a fixed magnetic field magnitude $\mu_0H=\SI{600}{\milli\tesla}$. For all distances ranging from $d=550-\SI{900}{\nano\meter}$ the amplitude $\Delta R_\mathrm{det}^\mathrm{el}$ increases with increasing temperature up to $\SI{200}{\kelvin}$ and starts to decrease again for higher temperatures. A very similar behavior was found in Ref.~\cite{Han2020}, which also studied the magnon spin transport in a hematite thin film of similar thickness in the easy-plane phase. 
We can explain the increase of $\Delta R_\mathrm{det}^\mathrm{el}$ with temperature by an increase of the magnon population in the AFI. This behavior has been captured in Eq.~\eqref{eq:Rnl}. The dominant quantities regarding the temperature dependence of the magnon spin signal are the interfacial spin conductivities $g_i$, $g_d$ and the susceptibility $\chi$, which increase similiarly with the number of magnons. However, at sufficiently high temperatures, the predominant effect appears to be magnon scattering, which leads to a decrease in their propagation length, as shown in Fig.~\ref{fig:fig2MITcomparison}(d).
Furthermore, we plot $\Delta R_\mathrm{det}^\mathrm{el}$ versus $d$ in Fig.~\ref{fig:fig2MITcomparison}(c), which allows us to extract the magnon diffusion length $l_\mathrm{m}$. The amplitude $\Delta R_\mathrm{det}^\mathrm{el}$ exponentially decreases with increasing distance as expected for diffusive magnon transport for $d>l_\mathrm{m}$~\cite{CornelissenMMR,Shan2016}. To extract $l_\mathrm{m}$, we can use the relation
\begin{equation}
\Delta R_\mathrm{det}^\mathrm{el}= \frac{C}{l_\mathrm{m}}\frac{\exp(d/l_\mathrm{m})}{1-\exp(2d/l_\mathrm{m})},
\label{eq:diffusionlength}
\end{equation}
where $C$ captures the distance-independent prefactors~\cite{CornelissenMMR}. The lines in Fig.~\ref{fig:fig2MITcomparison}(c) are fits to Eq.~\eqref{eq:diffusionlength} and the extracted $l_\mathrm{m}$ are shown in Fig.~\ref{fig:fig2MITcomparison}(d) as a function of temperature. We find a decrease of $l_\mathrm{m}$ with increasing temperature. While the temperature dependence slightly differs from the results in Ref.~\cite{Han2020}, which show an increase of the magnon diffusion length at low temperatures, the extracted magnon diffusion length $l_\mathrm{m}\approx \SI{0.15}{\micro\meter}$ at room temperature is in good agreement. In Ref.~\cite{Lebrun2020} a similiar magnon diffusion behavior is observed. In this analysis, we focused on magnetic fields much smaller then $H_\mathrm{c}$ and thus we expect that the magnon spin signal at the detector is dominated by the finite spin of the low-energy magnons. It is encouraging that all experimental studies obtain similar results indicating that the magnon transport is dominated by the properties of hematite and the injector/detector strip interface is only of minor concern. This further justifies our assumption of a weak coupling between injector/detector strip and the AFI.

\begin{figure}[tb]
	\begin{center}
		\includegraphics[width=85mm]{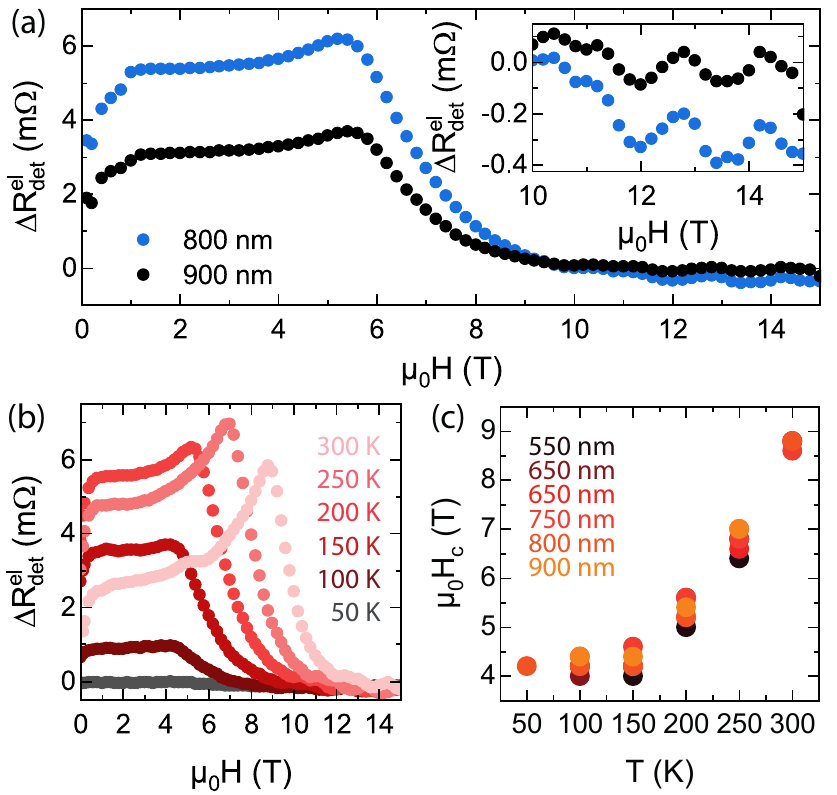}
		\caption{(a) Magnon spin signal amplitude $\Delta R_\mathrm{det}^\mathrm{el}$ of the electrically excited magnons as a function of the magnetic field magnitude $\mu_0H$, which is applied along $\hat{\pmb{y}}$ (cf. Fig.~\ref{fig:fig1thinhematite}), for center-to-center distances $d=\SI{800}{\nano\meter}$ and $d=\SI{900}{\nano\meter}$ between injector and detector on a $t_m=\SI{100}{\nano\meter}$ thick film measured at $T=\SI{200}{\kelvin}$. The inset shows a zoom-in on the oscillations for high magnetic fields. (b) $\Delta R_\mathrm{det}^\mathrm{el}$ plotted versus the magnetic field strength for a structure with a strip distance of $d=\SI{800}{\nano\meter}$ for different temperatures. (c) Compensation field $\mu_0H_\mathrm{c}$ as a function of $T$ extracted from measurements of structures with varying $d$.}
		\label{fig:fig3fielddependence100nm}
	\end{center}
\end{figure}

Last but not least, we extract the magnon spin signal amplitudes $\Delta R_\mathrm{det}^\mathrm{el}$ of the electrically excited magnons in the $\SI{100}{\nano\meter}$ thick film as a function of the  magnetic field magnitude $\mu_0H$ in the range $0$ to $\SI{15}{\tesla}$. The magnetic field is again applied along $\hat{\pmb{y}}$ and hence $\pmb{n}\parallel\pmb{\hat{z}}$ (as $\pmb{H}\perp\pmb{n}$). The data is exemplarily shown in Fig.~\ref{fig:fig3fielddependence100nm}(a) for two different injector-detector distances $d=\SI{800}{\nano\meter}$ and $\SI{900}{\nano\meter}$ for $T=\SI{200}{\kelvin}$. We find clear differences in the behavior of the $\SI{100}{\nano\meter}$ thick film compared to our thin film investigated in Sec.~\ref{sec:15nm_film}. Consistent with our previous observations, we clearly observe a peak in the positive magnon spin signal regime for both devices, here at $\mu_0H\approx\SI{5.5}{\tesla}$, which corresponds to the compensation field $\mu_0H_\mathrm{c}$. The peak position is independent of the center-to-center distance $d$, which originates from the fact that the pseudofield $\omega=0$ at $\mu_0H_\mathrm{c}$. The measured magnon spin signal amplitudes are about one order of magnitude larger compared to our thin films discussed above. Comparing our results to Eq.~\eqref{eq:Rnl}, one would expect a decreasing signal for increasing film thickness $t_\mathrm{m}$ if we assume that the other parameters remain unchanged. However, the dominant effect of the increasing film thickness is a higher density of magnonic states, which significantly enhances the various conductances involved. This explains the increase in the measured magnon spin signal.

For thicker films, a larger offset in the magnon spin signal $\Delta R_\mathrm{det}^\mathrm{el}$ for $\mu_0H<\SI{5.5}{\tesla}$ lets the Hanle peak appear smaller. 
We attribute the larger offset to the ordinary propagation of the finite-spin low-energy magnons. The observation of the typical magnon Hanle signal supports our model that the low-energy magnons do not contribute to the pseudospin precession, but rather simply to the finite-spin signal. In contrast, in our thin film the low-$k$ magnons are efficiently removed due to the standing wave-like situation along the film thickness.
In general, the low-energy magnon spin signal starts to diminish rapidly when the gap between the two magnon branches closes. Hence, we may associate an applied magnetic field with this point. The second magnetic field of relevance is the compensation field $\mu_0H_\mathrm{c}$. From our data, these two magnetic field values seem to differ slightly. At the same time, recent works~\cite{Shen2021,Ross2020} attribute the compensation field to the gap closing between the two magnon branches, which would suggest that both characteristic fields in our experiment overlap.

For $\mu_0H>\SI{5.5}{\tesla}$, $\Delta R_\mathrm{det}^\mathrm{el}$ decreases until it approaches zero.
Consistent with our model, we only observe a spin signal from low-energy magnons for small applied magnetic fields, as with increasing field strength their energy increases and they are then described via the pseudospin chemical potential. 
For further increasing field strength, the magnon spin signal starts to oscillate around $\Delta R_\mathrm{det}^\mathrm{el}=\SI{0}{\ohm}$ as clearly visible in the inset of Fig.~\ref{fig:fig3fielddependence100nm}(a), which shows a zoom-in of the data in the range $\mu_0H=10-\SI{15}{\tesla}$. This signal modulation does not show a clear dependence on the distance $d$ between injector and detector. At present, we do not have a convincing explanation for the physical origin of these oscillations and more work from both theory and experiment is required to better understand this interesting observation. 

Finally, we carried out measurements of the temperature dependence of the field-dependent magnon spin signal $\Delta R_\mathrm{det}^\mathrm{el}$. The data is shown in Fig.~\ref{fig:fig3fielddependence100nm}(b) for $d=\SI{800}{\nano\meter}$. For small applied magnetic fields, the offset finite-spin signal initially increases with increasing temperature until it starts to decrease again at a temperature of $T\approx\SI{200}{\kelvin}$. Moreover, the peak amplitude at $\mu_0H_\mathrm{c}$ also increases with temperature, however decreases again for temperatures above $T\approx\SI{250}{\kelvin}$. The oscillating behavior at higher fields is observed for all studied temperatures. The temperature dependence can be well explained via the two different contributions to the detector spin signal. At low temperatures, the low-energy magnons dominantly contribute to the spin transport, which is suppressed at high magnetic fields. At elevated temperatures, the higher energy magnons and the corresponding magnon Hanle signal is more and more contributing and the maximum in the spin signal at $\mu_0H_\mathrm{c}$ is better discernible from the low-energy magnon background.

For a quantitative analysis of  the compensation field, we extract $\mu_0H_\mathrm{c}$ as a function of temperature for different distances. The result is shown in Fig.~\ref{fig:fig3fielddependence100nm}(c). For each distance, we find a constant behavior in the temperature range from $50$ to $\SI{150}{\kelvin}$, while a significant increase is observed for larger temperatures. This is in perfect agreement with the qualitative behavior of thinner hematite films~\cite{Wimmer2020} and indicates that the compensation field $\mu_0H_\mathrm{c}$ follows the temperature dependence of the easy-plane anisotropy~\cite{Wimmer2020,Ross2020}.

\hspace{3.5cm}
%-----------------------------Conclusion----------------------------------------%
\section{Conclusion}

In conclusion, we systematically investigate the recently discovered magnon pseudospin dynamics and the associated magnon Hanle effect in hematite. This allows us to expand our previous description of the measured spin transport phenomena in terms of antiferromagnetic pseudospin dynamics. In our analysis we take into account both finite-spin contributions from low-energy magnons and the spin injection and detection process. This extended theoretical model is utilized to explain the influence of the effective dimensionality of the magnetic layer on the magnon spin signal. First, the influence of the injector-detector distance on the magnon spin signal is studied in a thin hematite film. We find an additional offset signal in the low field regime, which we could attribute to additional effects caused by an injector-detector spacing smaller than the magnon diffusion length. Second, we measured the magnon Hanle effect in a $\SI{100}{\nano\meter}$ thick hematite film. In contrast to the thinner films, peculiar changes are found. In particular, a pronounced offset signal in the low field regime is attributed to contributions from low-energy magnons and their finite-spin signal. Additionally, we observe an oscillating behavior of the magnon spin signal above $\SI{10}{\tesla}$, which shows no dependence on the electrode spacing distance.
Our work provides an important step towards the detailed understanding of magnonic pseudospin dynamics in antiferromagnetic systems and highlights the rich physics in antiferromagnetic magnonics.

%----------------- Acknowledgments --------------------------------------%
\section*{Acknowledgments}

We gratefully acknowledge financial support from the Deutsche Forschungsgemeinschaft (DFG, German Research Foundation) under Germany's Excellence Strategy -- EXC-2111 -- 390814868 and project AL2110/2-1 and the Spanish Ministry for Science and Innovation -- AEI Grant CEX2018-000805-M (through the ``Maria de Maeztu'' Programme for Units of Excellence in R\&D).

\end{document}